\RequirePackage{fix-cm}
\documentclass[twocolumn,epjc3]{svjour3}
\smartqed

\makeatletter
\@ifpackageloaded{natbib}{}{\usepackage{cite}}

\@ifpackageloaded{hyperref}{}{\usepackage[colorlinks,citecolor=blue,urlcolor=blue]{hyperref}}

\@ifundefined{oldSI}{%
\usepackage[separate-uncertainty,retain-explicit-plus,per-mode=symbol,product-units=power]{siunitx}
}{%
\usepackage[separate-uncertainty,retain-explicit-plus,per-mode=symbol,product-units=power]{siunitx}[=v2]
}
\makeatother

\DeclareSIUnit\cps{cps}
\DeclareSIUnit\electron{\mbox{$e^-$}}
\DeclareSIUnit\barg{barg}
\DeclareSIUnit\VoV{VoV}
\usepackage{lineno}
\usepackage{isotope}

\usepackage[utf8]{inputenc}
\usepackage[T1]{fontenc} 
\usepackage{appendix}
\usepackage{todonotes}
\usepackage{amsmath}
\usepackage{wasysym}
\usepackage{nicefrac}
\usepackage{tabularx}

\DeclareMathSymbol{\mdot}{\mathord}{symbols}{"01}

\title{SiPM cross-talk in liquid argon detectors}

\newcommand{\AQLNGS}{INFN Laboratori Nazionali del Gran Sasso, Assergi (AQ) 67100, Italy}
\newcommand{\AQGSSI}{Gran Sasso Science Institute, L'Aquila 67100, Italy}

\newcommand{\AstroCeNT}{AstroCeNT, Nicolaus Copernicus Astronomical Center of the Polish Academy of Sciences, 00-614 Warsaw, Poland}

\newcommand{\BOINFN}{INFN Bologna, Bologna 40126, Italy}

\newcommand{\Carleton}{Department of Physics, Carleton University, Ottawa, ON K1S 5B6, Canada}

\newcommand{\Princeton}{Physics Department, Princeton University, Princeton, NJ 08544, USA}

\newcommand{\RHUL}{Department of Physics, Royal Holloway University of London, Egham TW20 0EX, UK}
\newcommand{\RMTreINFN}{INFN Roma Tre, Roma 00146, Italy}

\newcommand{\RMUnoINFN}{INFN Sezione di Roma, Roma 00185, Italy}

\newcommand{\TNFBK}{Fondazione Bruno Kessler, Povo 38123, Italy}
\newcommand{\TNTIFPA}{Trento Institute for Fundamental Physics and Applications, Povo 38123, Italy}

\newcommand{\UCLA}{Physics and Astronomy Department, University of California, Los Angeles, CA 90095, USA}

\author{
  M.\,G.~Boulay\thanksref{LNGS, Carleton}
  \and V.~Camillo\thanksref{LNGS}
  \and N.~Canci\thanksref{LNGS}
  \and S.~Choudhary\thanksref{AstroCeNT}
  \and L.~Consiglio\thanksref{LNGS}
  \and A.~Flammini\thanksref{BOINFN}
  \and C.~Galbiati\thanksref{GSSI, Princeton} 
  \and C.~Ghiano\thanksref{LNGS}
  \and A.~Gola\thanksref{FBK, TIFPA}
  \and S.~Horikawa\thanksref{GSSI} 
  \and P.~Kachru\thanksref{GSSI} 
  \and I.~Kochanek\thanksref{LNGS} 
  \and K.~Kondo\thanksref{LNGS}
  \and G.~Korga\thanksref{RHUL, LNGS}
  \and M.~Ku\'zniak\thanksref{AstroCeNT}
  \and A.~Mazzi\thanksref{FBK, TIFPA}
  \and A.~Moharana\thanksref{GSSI}
  \and G.~Nieradka\thanksref{AstroCeNT}
  \and G.~Paternoster\thanksref{FBK, TIFPA}
  \and A.~Razeto\thanksref{LNGS}
  \and D.~Sablone\thanksref{LNGS}
  \and T.N.~Thorpe\thanksref{GSSI, UCLA}
  \and C.~T\"urko\u{g}lu\thanksref{AstroCeNT}
  \and H.~Wang\thanksref{UCLA}
  \and M.~Rescigno\thanksref{INFNRM1}
  \and S.~Sanfilippo\thanksref{INFNRM3}
}

\institute{
  \AQLNGS\thanksref{email} \label{LNGS}
  \and \Carleton \label{Carleton}
  \and \AstroCeNT \label{AstroCeNT}
  \and \BOINFN \label{BOINFN}
  \and \AQGSSI \label{GSSI}
  \and \Princeton \label{Princeton}
  \and \TNFBK \label{FBK}
  \and \TNTIFPA \label{TIFPA}
  \and \RHUL \label{RHUL}
  \and \UCLA \label{UCLA}
  \and \RMUnoINFN \label{INFNRM1}
  \and \RMTreINFN \label{INFNRM3}
}

\thankstext{email}{Corresponding address: sarlabb7@lngs.infn.it}

\journalname{Eur. Phys. J. C}
\begin{document}
\maketitle

\begin{abstract}
SiPM-based readouts are becoming the standard for light detection in particle detectors given their superior resolution and ease of use with respect to vacuum tube photo-multipliers. However, the contributions of detection noise such as the dark rate, cross-talk, and after-pulsing may impact significantly their performance.  In this work, we present the development of highly reflective single-phase argon chambers capable of light yields up to \num{32} photo-electrons per keV, with roughly \num{12} being primary photo-electrons generated by the argon scintillation, while the rest are accounted by optical cross-talk.  Furthermore, the presence of compound processes results in a generalized Fano factor larger than \num{2} already at an over-voltage of \SI{5}{\volt}.  Finally, we present a parametrization of the optical cross-talk for the FBK NUV-HD-Cryo SiPMs at \SI{87}{\kelvin} that can be extended to future detectors with tailored optical simulations.

\end{abstract}

\section{Introduction}
In 1955 it was observed that a silicon junction emits light when a bias is applied across it~\cite{Newman}. More recently, a number of works have documented the emission of light by SiPMs during the avalanche process~\cite{hama-emission}.  Internal cross-talk (iCT) is when such a photon remains confined within the source SiPM and generates another avalanche in a neighbouring cell.  External cross-talk (eCT) is when the generated photon escapes from the silicon bulk reaching another SiPM array in the experimental setup.  Finally, feedback cross-talk (fCT) is when the photon undergoes reflection and is reabsorbed by the same SiPM array which emitted it.  We define optical cross-talk (oCT) as the envelope which includes the three of these effects.  As we will describe, optical cross-talk generates a compound process that leads to the amplification of the initial signal with a gain defined by $G=\nicefrac{1}{(1-\lambda_\textrm{oCT})}$ (where $\lambda_\textrm{oCT} \ll 1$ is the average number of secondary avalanches following any avalanche in the process). 
Unfortunately, such processes are subject to fluctuations that affect the resolution of the measurement.  This is addressed in Section~\ref{sec:fano}, where we quantify a generalized Fano factor that is significantly larger than unity.

Many particle detectors are designed to collect very faint light signals in chambers that host several thousand of photo-detectors, such as Borexino and Super-Kamiokande~\cite{borexino, superk}. In these conditions, oCT between the photo-detectors can have a large impact on the physics results of the experiments.  Argon is of particular interest as an active detector medium because of its high scintillation yield.  Multiple large particle detectors have made, or will make, use of liquid argon (LAr)~\cite{icarus, dune, deap3600, ds20k}.  In this work we present the development of two high efficiency LAr chambers which were operated at \SI{87}{\kelvin} to study the scintillation light produced by the interactions with calibration sources.  We then present the deconvolution of the oCT into the individual contributions. 

\section{Experimental Setups}
The detectors were installed in a sealed dewar inside a container filled with roughly \SI{4}{\liter} of high purity LAr, within the STAR facility~\cite{2pac}.  The system consists of a re-circulation loop capable of a volumetric flow of \SI{5}{sl\per\minute}, and provides continuously purified argon via a getter (SAES PS4-MT3).

Two radioactive sources are used.  The meta-stable isotope, \isotope[83\rm m]Kr, with an activity of $\mathcal{O}$(10) Bq can be injected into re-circulation loop.  \isotope[83\rm m]Kr has been used in previous direct dark matter experiments \cite{ds50first} as it is not filtered by the getters, and provides a calibration line at \SI{41.5}{\kilo\eV}.  \isotope[241]Am can be attached to the external wall of the dewar, providing \SI{59.5}{\kilo\eV} gamma-rays within the active volume of the detector at a rate of several Hertz.

\subsection{Cubic Chamber}
\begin{figure}[tb]
\centering
\includegraphics[width=0.48\textwidth, ]{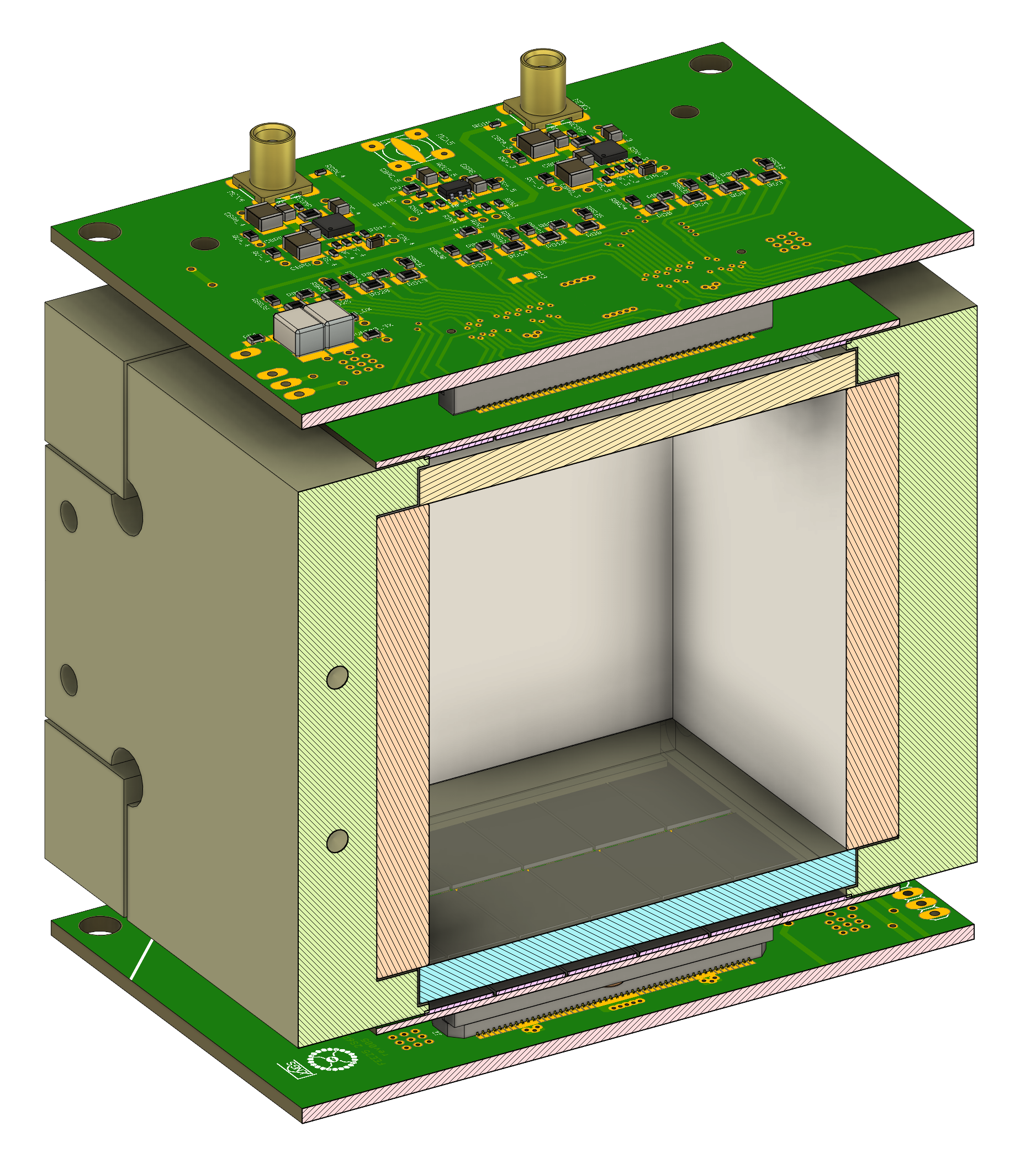}
\caption{Drawing of the cubic chamber with inner dimensions \SI[product-units=power]{50 x 50 x 50.8}{\mm} (l x w x h).  High efficiency reflectors are installed on the chamber walls.  Two \SI{24}{\square\cm} SiPM tiles, with their readout boards, are facing UV-grade fused silica windows which are installed on the top and the bottom.  All the inner surfaces are evaporated with TPB.}
\label{fig:cubic_chamber}
\end{figure}

Figure~\ref{fig:cubic_chamber} depicts the cubic LAr chamber.  The chamber consists of four identical walls machined from poly-etheretherketone (PEEK) which can host different reflectors.  The results reported here refer to an enhanced specular reflector (Vikuiti ESR) from the 3M company. The top and bottom windows are $1\lambda$ fused silica with dimensions of \SI[product-units=power]{50 x 50 x 4}{\mm}. All internal surfaces are coated with tetraphenyl butadiene (TPB) for wavelength shifting of the scintillation photons (\SI{128}{\nano\meter}) to the visible range of the spectrum.

The visible photons are detected by two SiPM arrays (tiles) installed on the top and bottom of the detector chamber.  Each tile is made up of twenty-four SiPMs (summed into quadrants) bonded to a FR4 PCB with a cryo-grade epoxy~\cite{iza-bari}.  The SiPMs from the FBK NUV-HD-Cryo family have a surface area of \SI[product-units=power]{7.9 x 11.7}{\mm}, with a cell size of \SI{30}{\micro\meter} and a quenching resistor of \SI{5}{\mega\ohm} at \SI{87}{\kelvin}~\cite{nuv-hd-cryo}.  The fill-factor of the tiles is \SI{90}{\percent}, where most of the dead space is reserved for the landing pad of the wire bonding.

\subsection{Cylindrical Chamber}
A cylindrical chamber with inner dimensions \SI[product-units=power]{46 x 50}{\mm} ($\diameter$ x h) was instrumented with the same photo-detectors as used in the cubic chamber. %%, Figure~\ref{fig:cyl_chamber}.  
The cylinder is made of acrylic, and internally lined with TPB-coated 3M reflector foil.  In front of the SiPM surface, two \SI{1}{\milli\meter} thick TPB-coated fused silica windows are installed.  The cylindrical chamber is useful for verifying the consistency of the results and of the models developed for the cubic chamber.

\subsection{Data Acquisition}
\label{sec:daq}

Each tile is connected to a readout board where the signals from the SiPMs are amplified by four cryo-grade low-noise trans-impedance amplifiers \cite{tile}.  In this configuration, we achieve a signal to noise ratio (SNR) larger than \num{7} for the unfiltered signal, and an SNR larger than \num{30} for the charge in \SI{1}{\micro\second} at an over-voltage (OV) of \SI{5}{\volt}.

Waveforms are acquired with a V1720 CAEN digitizer, with each tile quadrant connected to a channel.  A copy of the signal from the SiPMs is sent to a set of NIM discriminators that form a trigger logic.  The trigger can be configured to act on a single tile or on both, depending on the experimental conditions.  The trigger threshold is set significantly below the region of interest.

Data was acquired up to \SI{9.5}{OV} with \isotope[241]Am, \isotope[83\rm m]Kr, and with no radioactive source present.  Resulting background subtracted data sets of the calibration energy peaks are then obtained.

\begin{figure}[tb]
\centering
\includegraphics[width=0.48\textwidth, ]{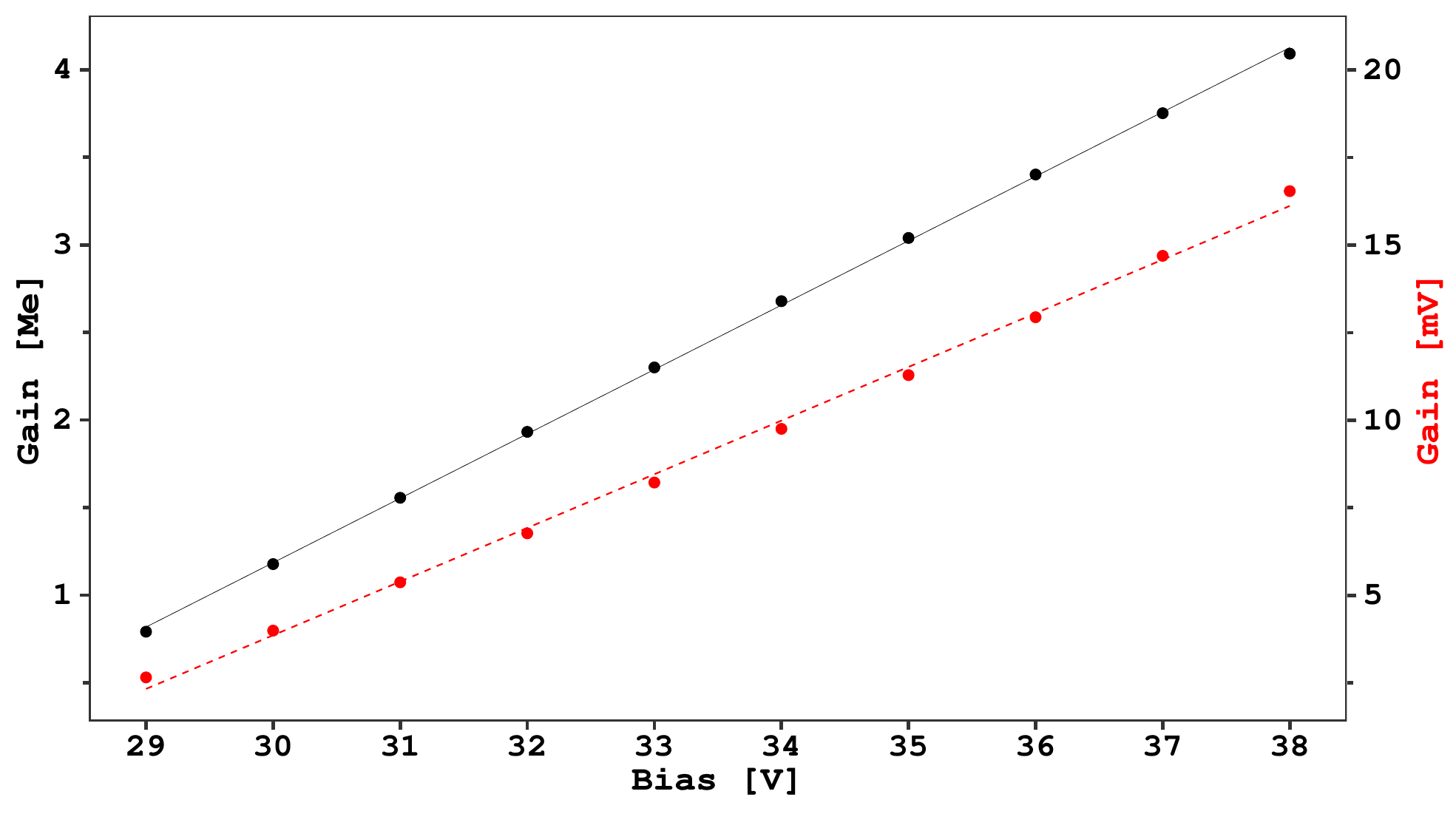}
\caption{Gain in $e^-$ (black) and in peak amplitude (red) for the SiPMs in use as a function of the applied bias. The solid and dashed lines correspond to linear regressions, which describe the experimental data to better than \SI{1}{\percent} and \SI{4}{\percent} for charge and amplitude, respectively.}
\label{fig:gain}
\end{figure}

A laser pulse is delivered in the chambers to monitor the behaviour of the photo-detectors: in Figure~\ref{fig:gain} we report the charge gain and the peak amplitude for one of the SiPMs in use.  Both quantities are described to better than a few percent by linear models (as a function of over-voltage). The break-down voltages are measured as $V_{bd}^C$= \SI{26.8\pm0.1}{\volt} and $V_{bd}^A$=\SI{27.5\pm0.1}{\volt} for charge and amplitude, respectively.  From here, the over-voltage will be relative to $V_{bd}^A$.

\section{Data Analysis}
The data analysis involves integrating the normalized waveforms over a gate of \SI{7}{\micro\second} following the NIM trigger to obtain the photo-electron (PE) spectrum.  Over this time scale, \SI{99.5}{\percent} of the Ar scintillation light is emitted~\cite{DEAP:pulseshape}, and delays introduced by the absorption and re-emission of photons in the wavelength shifter (WLS) and their optical path length inside the detector have little effect.  The normalized waveforms are scaled by the gain of the photo-detectors and the baseline, which is calculated in the pre-trigger region, is removed.  The mean number of photo-electrons observed by both photo-detectors is extracted by fitting the calibration peak data with a Gaussian model~\cite{2pac}.

\begin{figure}[tb]
\centering
\includegraphics[width=\columnwidth, ]{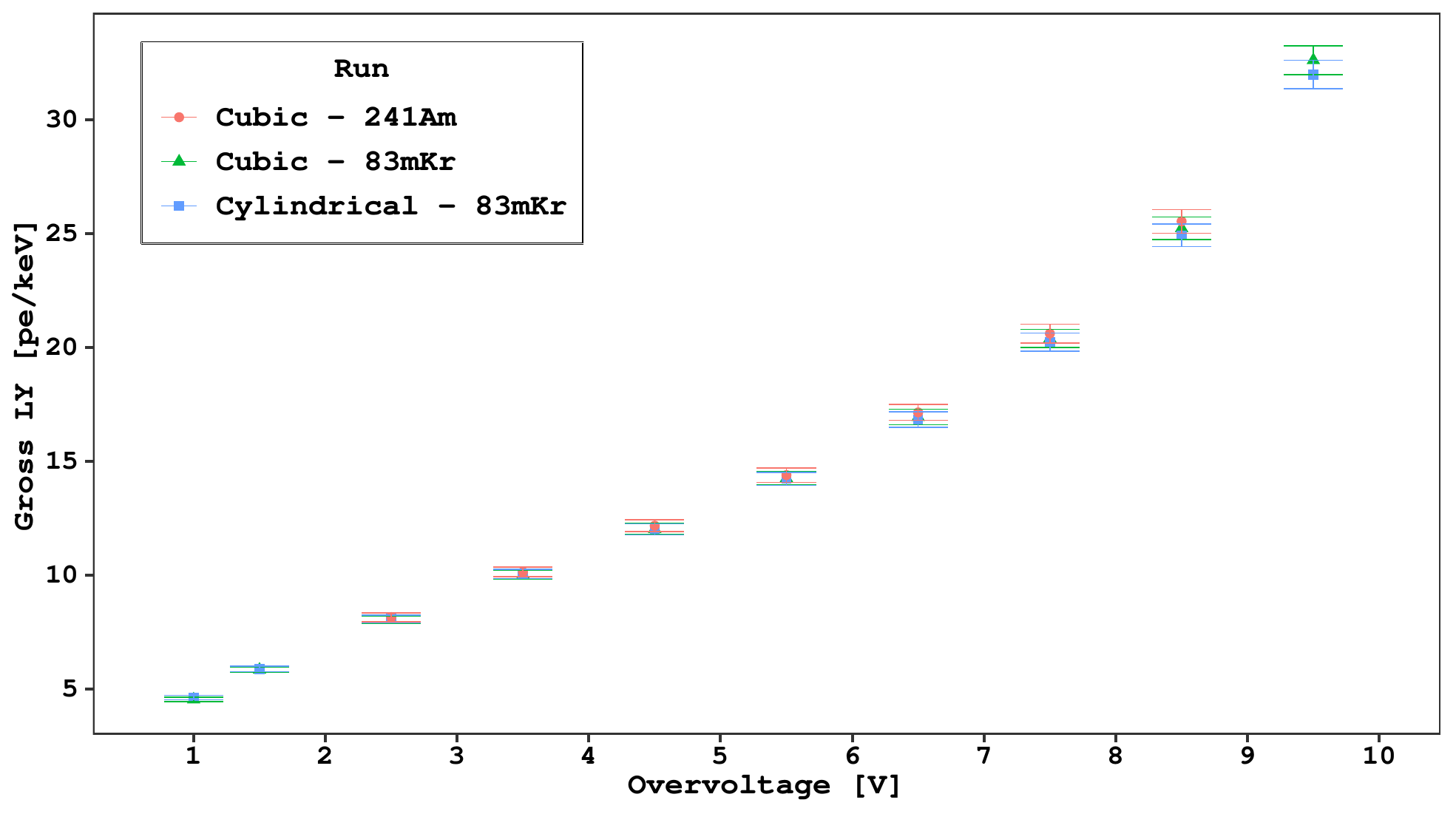}
\caption{Gross light yield (without correcting for correlated noise) measured at different SiPM over-voltages for the cubic chamber using  \isotope[83\rm m]Kr (green triangles) and \isotope[241]Am (red circles) radioactive sources, and for the cylindrical chamber using the \isotope[83\rm m]Kr (blue squares) radioactive source.}
\label{fig:LYG}
\end{figure}

The gross light yield, LY$_G^\textrm{OV}$, is the ratio of the number of detected photo-electrons to the energy deposited within the medium by the radioactive source.  The gross energy resolution, $\sigma_G^\textrm{OV}/E$, is determined by the standard deviation divided by the mean of the fitted Gaussian model.  Figure~\ref{fig:LYG} shows LY$_G^\textrm{OV}$ versus over-voltage for the cubical and the cylindrical chambers, where similar values are obtained for both radioactive sources.

\subsection{Internal Cross-talk (iCT) and After-pulsing (AP)}
\label{sec:ict}

\begin{figure}[tb]
\centering
\includegraphics[width=\columnwidth]{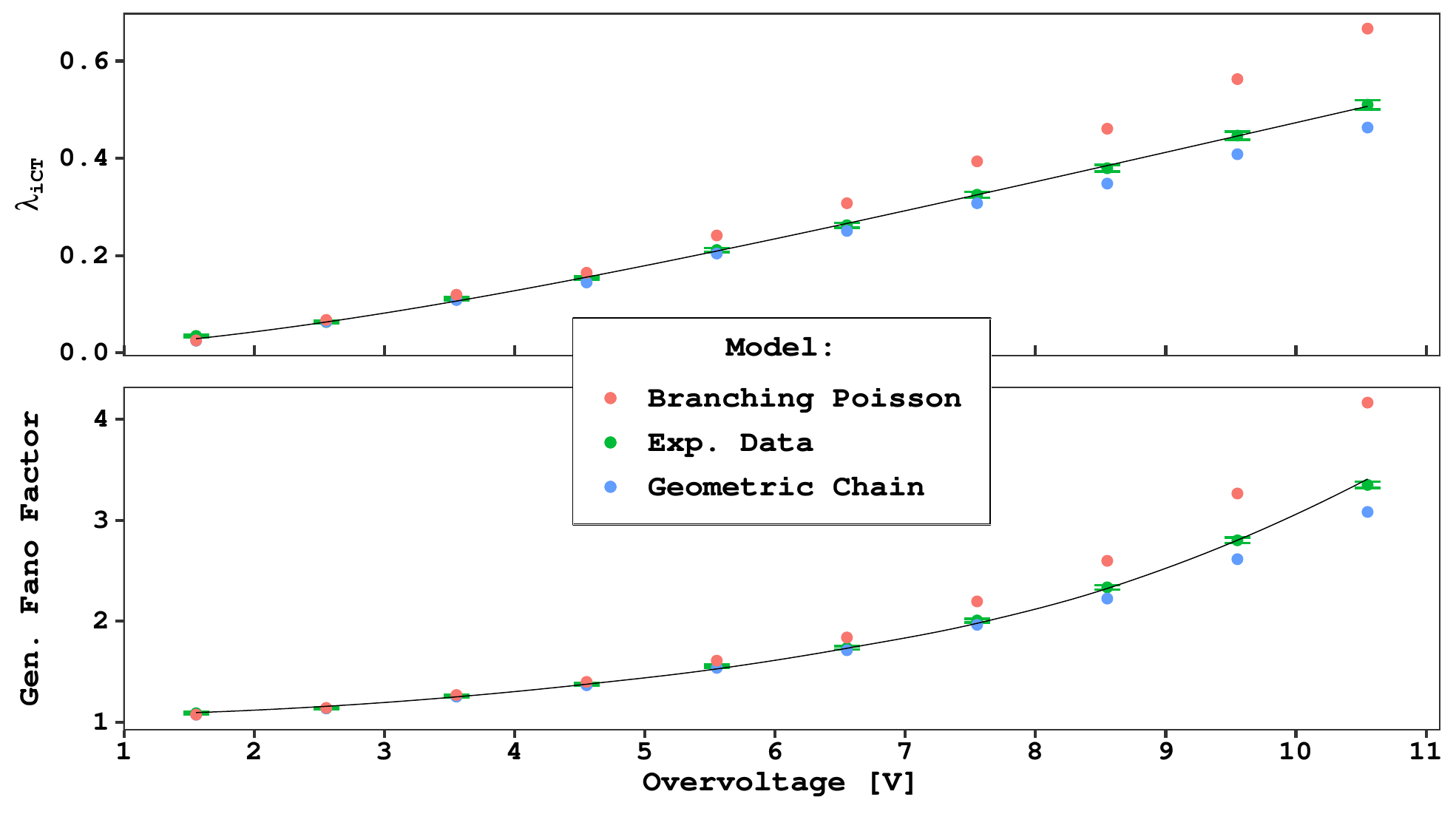}
\caption{iCT analysis for the SiPMs used.  The experimental data points (green) are extracted from raw data with Equation~\ref{eqn:ict:data}, and the solid black lines represent best fits to the data using Equations~\ref{eqn:ict:model} (top) and \ref{eqn:ict:fano} (bottom).}
\label{fig:iCT}
\end{figure}

Figure~\ref{fig:iCT} reports two figures of merit for the iCT for the SiPMs in use, measured at  \SI{77}{\kelvin}.  These measurements are performed with single SiPMs from the same lot as the photo-detectors of the cylindrical and cubic chambers.  The SiPMs are exposed to laser pulses in a stable, low-noise environment.  During analysis, up to twenty photo-electron peaks were identified.

In this work, we assume that the cross-talk photons are emitted independently in a recursive process converging at $\nicefrac{\epsilon}{(1-\lambda_\textrm{iCT})}$, where $\lambda_\textrm{iCT}$ is the average number of secondary avalanches following any avalanche in the process. As in Ref.~\cite{vinogradov-coupound}, the model is valid for $\lambda_\textrm{iCT} \ll 1$.

Using the charge spectra (a.k.a. ``finger plots''), the relative population of each peak, $\mathcal{R}^\textrm{OV}_n$ (relative to $n$ photo-electrons), is extracted.  We can assume that the detected laser photons follow a Poissonian distribution with a mean value, $\epsilon$ (kept much smaller than unity). The following quantities are defined:
\begin{align}\begin{split}
\lambda_\textrm{iCT}^\textrm{OV} & = 1 + \frac{ln(\mathcal{R}^\textrm{OV}_0)}{ \langle n\rangle} \\
\mathcal{F}_\textrm{iCT}^\textrm{OV} & = \frac{\textrm{Var}\left[n\right]} {\langle n\rangle}
\label{eqn:ict:data}
\end{split}\end{align}
where $\epsilon = -ln(R^\textrm{OV}_0)$, $\textrm{Var}[n]$ and $\langle n \rangle$ define the variance and the mean of the photo-electron peak distribution ($\mathcal{R}^\textrm{OV}_n$), respectively.
Equations~\ref{eqn:ict:data} provide the maximum likelihood estimate for $\lambda_\textrm{iCT}$ and the generalized Fano factor ($\mathcal{F}_\textrm{iCT}$) with the experimental data, green points in Figure~\ref{fig:iCT}. 

The generalized Fano factor ($\mathcal{F}_\textrm{iCT}$) is defined as a simple variance-to-mean ratio 
and quantifies the deviation from a Poisson model. 
If the primary photo-electrons follow a Poisson distribution, this generalized Fano factor is related to the 
Excess Noise Factor ($ENF$) as $\mathcal{F} = G\cdot ENF$, 
where $G$ is the amplification. In our set-up, such amplification is provided by the iCT, therefore  $G=\nicefrac{1}{(1-\lambda_\textrm{iCT})}$.

We model the behaviour of $\lambda_\textrm{iCT}^\textrm{OV}$ and of $\mathcal{F}_\textrm{iCT}^\textrm{OV}$ with the following equations:
\begin{align}
\lambda_\textrm{iCT}(V)& = \xi_\textrm{iCT} \mdot (V-V_{bd}^C) \mdot P^h_T(V-V_{bd}^A)
\label{eqn:ict:model}\\
\mathcal{F}_\textrm{iCT}(V)& = \delta \mdot (1-\lambda_\textrm{iCT}(V))^\alpha
\label{eqn:ict:fano}
\end{align}
where the two break-down voltages ($V_{bd}^A$ and $V_{bd}^C$) are determined by the laser data analysis described in Section~\ref{sec:daq}. Equation~\ref{eqn:ict:model} follows the parametrization introduced in Ref.~\cite{montarulli}, where $\xi_\textrm{iCT}$ represents the acceptance for the iCT process. Equation~\ref{eqn:ict:fano} describes the generalized Fano factor with an effective model with two free parameters $\alpha$ and $\delta$. 

The triggering probabilities for hole and electron initiated avalanches ($P^h_T$ and $P^e_T$) are parametrized with exponential dependencies as in Ref.~\cite{mcintyre}. 
We define:
\begin{align}
\begin{split}
P^h_T(\Delta\textrm{V})& = 1 - e^\frac{\Delta\textrm{V}}{-V_h} \\
P^e_T(\Delta\textrm{V})& = 1 - e^\frac{\Delta\textrm{V}}{-V_e} 
\end{split}
\label{eqn:pt}
\end{align}
where $V_h$ and $V_e$ describe the temperature dependent mean energy required by a drifting carrier to extract charge with inelastic scattering. We found that the triggering probabilities for our data are better described in terms of $\Delta V=V-V_{bd}^A$ as in Ref.~\cite{zappala}, as opposed to Ref.~\cite{otte} that defines $\Delta V_\textrm{Otte}=V-V_{bd}^C$ or to Ref.~\cite{dinu} that defines $\Delta V_\textrm{Dinu} = V - V_{bd}^{IV}$ (where $V_{bd}^{IV}$ is the break-down voltage defined by the I-V curve).

Equation~\ref{eqn:ict:model} describes the iCT process in terms of emission and trigger probabilities.  The iCT photons are peaked in the red - infrared region~\cite{hama-emission}, where the detection for the NUV SiPMs is carried by holes. Therefore in Equation~\ref{eqn:ict:model} only the hole triggering probability is used.  
The emission probability is considered to be proportional to the total number of carriers extracted in the avalanche process, which is directly proportional to $V-V_{bd}^C$ (at better than \SI{1}{\percent}, see Figure~\ref{fig:gain}).

The fits converge to: 
$V_h =\SI{5.4\pm0.3}{\volt}$ and $\xi_\textrm{iCT}=\SI{53\pm1}{\per\kilo\volt}$ with  $\chi^2/\textrm{d.o.f.} = 9/8$, and 
$\alpha =\num{-1.68\pm0.01}$ and $\delta=\num{1.031\pm0.008}$ with $\chi^2/\textrm{d.o.f.} = 13/8.$ 
Figure~\ref{fig:iCT} (solid black lines) reports the prediction of the model using these parameters.
\\

Figure~\ref{fig:iCT} also shows the  {\it branching Poisson} (BP) and {\it geometric chain} (GC) models  as described by Vinogradov~\cite{vinogradov-analytical, vinogradov-coupound}. 
The top panel shows $\lambda^\textrm{GC}_\textrm{iCT}$ and $\lambda^\textrm{BP}_\textrm{iCT}$ resulting from the fit of $\mathcal{R}^\textrm{OV}_n$ with the corresponding model. 
The bottom panel shows the generalized Fano factor, which is calculated for GC and BP as 
$\mathcal{F}^\textrm{GC}_\textrm{iCT}=\nicefrac{(1+\lambda_\textrm{iCT})}{(1-\lambda_\textrm{iCT})}$ 
and $\mathcal{F}^\textrm{BP}_\textrm{iCT}=\nicefrac{1}{(1-\lambda_\textrm{iCT})^2}$, respectively. 

Both models depart from data by more than \SI{10}{\percent} at the highest over-voltages.  
Our data is better modeled by the sum of two binomial processes with probabilities $85\% \mdot \lambda_\textrm{iCT}$ and $15\% \mdot \lambda_\textrm{iCT}$.  
In this way each avalanche can generate 0, 1, or 2 photo-electrons in neighbouring cells, where a similar process occurs recursively.  The accuracy of this \textit{effective} model is better than \SI{0.2}{\percent} for the resulting mean number of iCT photo-electrons and their generalized Fano factors. \\

After-pulsing (AP) was studied in Ref.~\cite{2pac}, and for the over-voltages used here it does not exceed \SI{10}{\percent}.  The primary dark rate (DCR) does not exceed \SI{20}{\cps} per photo-sensor in cryogenic conditions.  As a first approximation, we do not consider these quantities.

\subsection{External cross-talk (eCT)}
\label{sec:ect}

The eCT contribution is measured directly by scanning over-voltages with one photo-detector (source), while holding the other (target) at a constant value.

\begin{figure}[tb]
\centering
\includegraphics[width=\columnwidth]{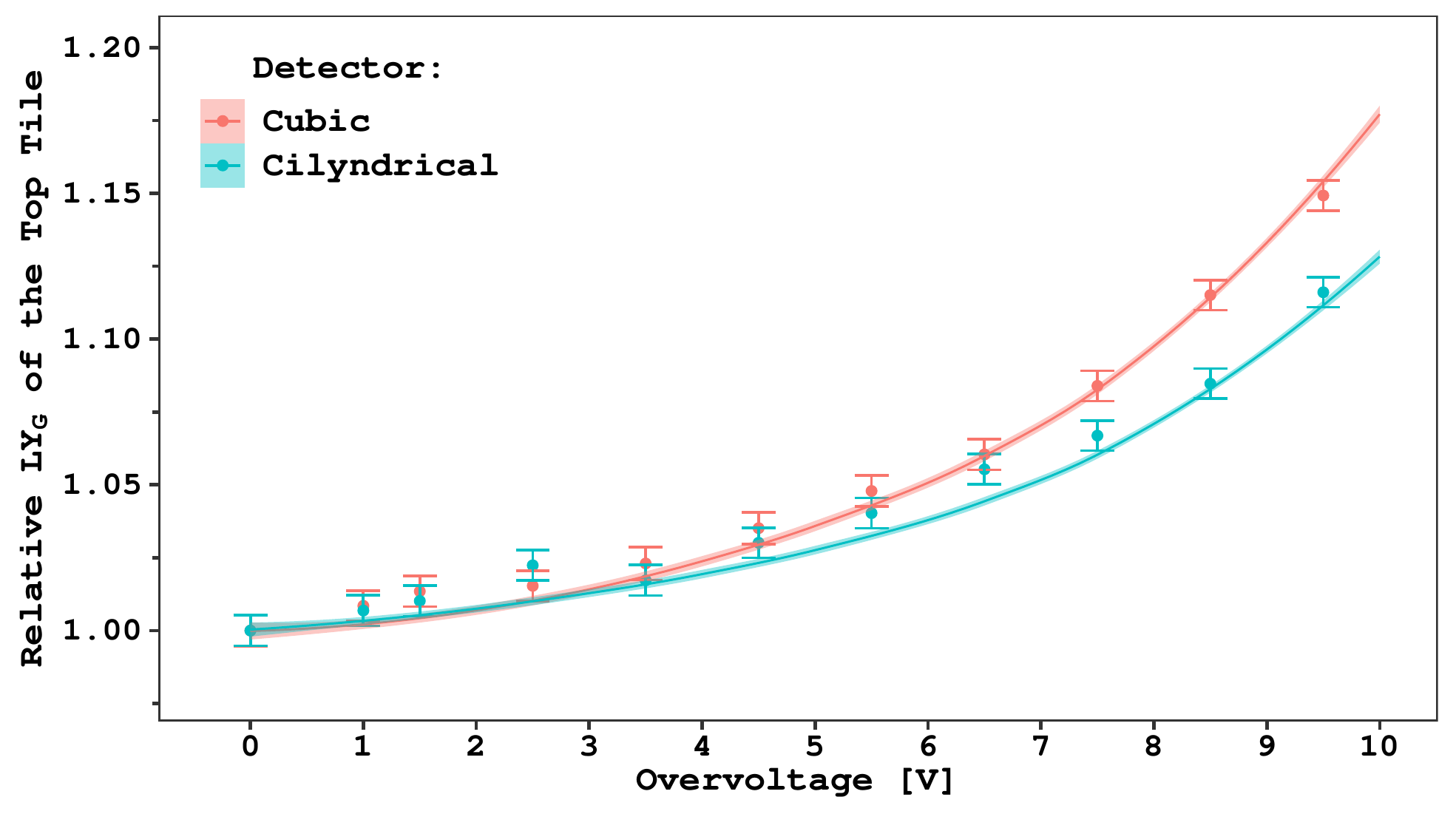}
\caption{External cross-talk versus over-voltage.  This shows the relative amount of light seen by the top photo-detector (target at \SI{8.5}{OV}) for different over-voltages set on the bottom photo-detector (source). The light yield is scaled to \SI{0}{OV}. The solid lines represent the toy Monte Carlo fit to the data described in Section~\ref{sec:tmc}.  The small confidence intervals around the lines come from the statistical fluctuation of roughly $10^9$ photons in the simulation, not from the uncertainty of the fit parameters.}
\label{fig:eCT}
\end{figure}

Figure~\ref{fig:eCT} shows the relative increase of LY$_G$ measured by the target versus the over-voltage of the source.  In this case the calibration peak can no longer be modeled as a Gaussian.  The LY$_G$ of the target detector is estimated by the mean of the distribution with the \isotope[83\rm m]Kr radioactive source after background subtraction.

With equal tile biases (Figure~\ref{fig:LYG}) similar values of LY$_G^\textrm{OV}$ were obtained with both chambers; however, the eCT component was found to be lower for the cylindrical chamber.  One explanation is the circular cross-section of the chamber being roughly two-thirds the area of the SiPM array, with a large fraction of the SiPM surface facing the aluminum support frame, which may be resulting in a larger (smaller) fraction of fCT (eCT). 

\begin{table*}[t]
\centering
\begin{tabular}{|l|r|r|r|r|r|r|c|}
\hline
 & \multicolumn{4}{c|}{\bf Cubic} & \multicolumn{2}{c|}{\bf Cylindric} & {\bf Units} \\
\cline{1-7} 
Bias & Asymmetric & \multicolumn{3}{c|}{Symmetric} & Asymmetric & Symmetric & \\
\cline{1-7} 
Algorithm  & \multicolumn{2}{c|}{tMC} & \multicolumn{1}{c|}{Analytical} & \multicolumn{1}{c|}{Global} & \multicolumn{1}{c|}{tMC} & \multicolumn{1}{c|}{Analytical} &  \\
\hline
$\bar{n}^\textrm{pe}$ & \num{13.0\pm0.5} & \num{12.1\pm0.3} & \num{13.0\pm.9} & \num{12.9\pm0.9} & \num{13.0\pm0.7} & \num{12\pm2} & pe/keV\\

$\zeta$ & \num{.34\pm0.05} & \num{.34\pm0.10} & \num{.34\pm.08} & \num{.37\pm0.08} & \num{.34\pm0.07} & \num{.45\pm.34} & -\\ 

$V_h$ & 5.4 & 5.4 & \num{5.4\pm0.3} & \num{5.6\pm0.3} & 5.4 & \num{5.3\pm0.3} & V\\

$V_e$ & \num{1.0\pm0.1} & \num{1.0\pm.1} & \num{0.9\pm0.2} & \num{1.0\pm0.2} & 1 & \num{1.1\pm0.6} & V \\

$\xi_\textrm{iCT}$ & 53 & 53 & 53 & 53 & 53 & 53 & kV$^{-1}$ \\

$\xi_\textrm{fCT}$ & \num{15\pm1} & 15 & 15 & 15 & \num{17\pm1} & 17 & kV$^{-1}$\\

$\xi_\textrm{eCT}$ & \num{7\pm1} & \num{7\pm1} & \num{8\pm2} & \num{10.2\pm1.5} & \num{5\pm1} & \num{8\pm5} & kV$^{-1}$\\

$\alpha$ & &  &  & \num{-1.71\pm0.07} & & & -\\

$\delta^*$ & &  &  & \num{1.16\pm0.01} & & & - \\

$\chi^2/n.d.f.$ & 4 / 5 & 3 / 6  & 1 / 5 & 13 / 13 & 11 / 6 & 2 / 5 & -\\
\hline
\end{tabular}
\caption{Results of the fits to the data for the cubic and the cylindrical chambers.  Asymmetric biasing refers to the eCT scan where one photo-detector (target) is kept at constant bias and the other (source) is scanned over the range \SIrange{0}{10}{OV}, see Section~\ref{sec:ect}.  Symmetric biasing corresponds to normal operation where both photo-detectors are held at the same bias.  The tMC, the analytical model, and the global fits are reported for the cubic chamber.  The parameters without errors are fixed while $V_h$ (when not fixed) has a penalty in $\chi^2$ at \SI{5.4\pm0.3}{\volt} (see text).}
\label{tab:ly:fit}
\end{table*}

\subsection{Toy Monte Carlo Simulation (tMC)}
\label{sec:tmc}
To quantify the different oCT components, a toy Monte Carlo simulation (tMC) was developed. The tMC models the detection of primary photons and the intertwined cross-talk photo-electrons as a recursive binomial process, similar to the geometric compound process from Vinogradov~\cite{vinogradov-coupound} described in Section~\ref{sec:ict}.  Photon tracking is not part of the tMC, and only the overall acceptances ($\xi_x$) for each sub-process (primary photo-electrons, iCT, fCT and eCT) are used.

\begin{figure}[tb]
\centering
\includegraphics[width=\columnwidth]{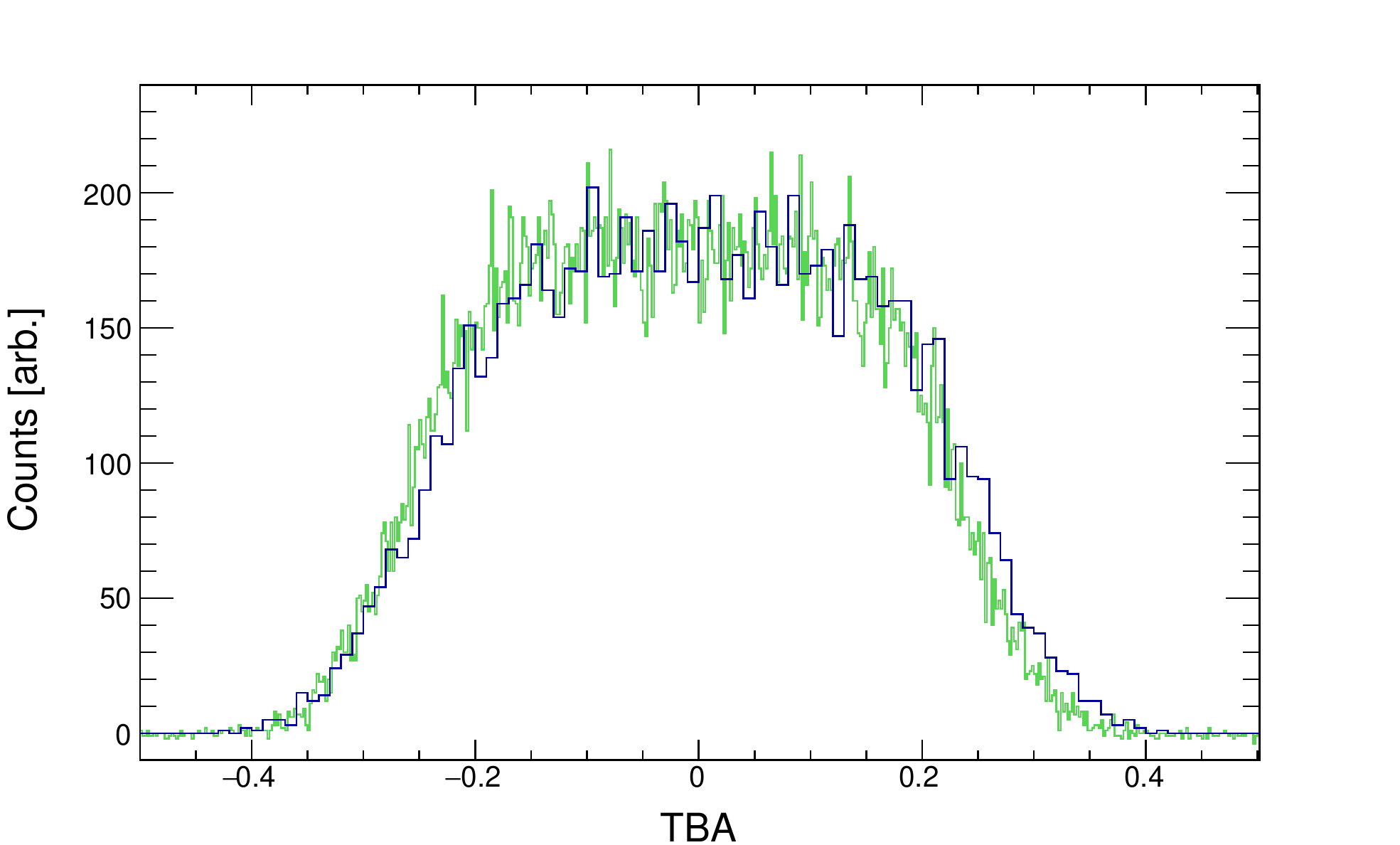}
\caption{Top-Bottom Asymmetry (TBA) using the \isotope[83\rm m]Kr radioactive source in the cubic chamber.  Experimental data measured at \SI{8.5}{OV} (green) with an asymmetry below \SI{1.5}{\percent} and simulated data (blue), which is described in Section~\ref{sec:simulation}.}
\label{fig:asym}
\end{figure}

The tMC assumes a symmetric detector, which is justified by Figure~\ref{fig:asym} where the top-bottom asymmetry (TBA) is shown for the cubic chamber.  The TBA is defined, at the event level, as the difference in photo-electrons seen by the top and bottom tiles normalized to the total number of collected photo-electrons.  Under this hypothesis the following quantities are defined:
\begin{align}
&P_\textrm{PDE}(V) =~ \zeta \mdot P^e_T(V-V_{bd}^A)  + (1-\zeta) \mdot P^h_T(V-V_{bd}^A)
\label{eqn:tmc:a}\\
& \lambda_\textrm{xCT}(V_S, V_T)  = \xi_x \mdot (V_S-V_{bd}^C) \mdot P^h_T(V_T-V_{bd}^A)
\label{eqn:tmc:b}
\end{align}

Equation~\ref{eqn:tmc:a} defines the probability of detecting a primary photo-electron as a function of the probability of creating an electron ($\zeta$) or hole ($1-\zeta$) (see Ref.~\cite{zappala}) with their relative triggering probabilities, defined in Equation~\ref{eqn:pt}.  $V_h$ is set to the value extracted by fitting Equation~\ref{eqn:ict:model} to the data in Figure~\ref{fig:iCT}.  
Equation~\ref{eqn:tmc:b} generalizes Equation~\ref{eqn:ict:model}, modeling the probability of emission and detection of cross-talk photo-electrons as a function of the bias of the source and of the target photo-detectors.  For iCT and fCT, $V_S$ and $V_T$ coincide.  The acceptance parameters are $\xi_\textrm{fCT}$, $\xi_\textrm{eCT}$ and $\xi_\textrm{iCT}$, the last of which is set to the value returned from Equation~\ref{eqn:ict:model}.

The $\chi^2$ minimization of the five free parameters against the data of the over-voltage scan for the cubic chamber (Figure~\ref{fig:eCT}) leads to an average accuracy better than 1.5 photo-electrons over about 500 photo-electrons. The resulting parameters for both the cubic and the cylindrical chambers are reported in Table~\ref{tab:ly:fit}.

\subsection{Analytical model}
An analytical model was independently developed based on simple mathematical assumptions; the tMC helps validate this model.  The basic assumption of the model is that the iCT recursive process introduces a photo-electron gain ($\mu$) that can be generalized in presence of mutually interacting iCT and fCT (described by $\lambda_\textrm{fCT}$ in analogy to $\lambda_\textrm{iCT}$) as follows:
\begin{align}
\mu(V)=\frac{1}{1-(\lambda_\textrm{iCT}(V)+\lambda_\textrm{fCT}(V))}
\label{eqn:compound:mu}
\end{align}
that is valid only for $\lambda_\textrm{iCT}+\lambda_\textrm{fCT} \ll 1$.

Equation~\ref{eqn:compound:fct} represents the mean number of photo-electrons in the presence of iCT and fCT for an event of energy, E. 
\begin{align}
N^\textrm{pe}(V) 
= \bar{n}^\textrm{pe}\mdot E\mdot P_\textrm{PDE}(V) \mdot \mu(V)
\label{eqn:compound:fct}
\end{align}
where the parameter  $\bar{n}^\textrm{pe}$ indicates the asymptotic net light yield and $P_\textrm{PDE}$ is defined in Equation~\ref{eqn:tmc:a}.

With two SiPM arrays and in presence of external cross-talk, we can define 
$\lambda_{\textrm{eCT}_{12}}$ ($\lambda_{\textrm{eCT}_{21}}$) as the number of avalanches in detector 2 (1) caused by a photo-electron in detector 1 (2). Equation~\ref{eqn:compound:fct} can be extended to:
\begin{align}
\begin{split}
N^\textrm{pe}_1(V_1, V_2) =&~ \mu_1(V_1) \mdot (\bar{n}_1^\textrm{pe} \mdot E\mdot P_{\textrm{PDE}_1}(V_1) + \\
&+ N^\textrm{pe}_2(V_1, V_2) \mdot \lambda_{\textrm{eCT}_{21}}(V_2, V_1)) \\
N^\textrm{pe}_2(V_1, V_2) =&~ \mu_2(V_2) \mdot (\bar{n}_2^\textrm{pe} \mdot E\mdot P_{\textrm{PDE}_2}(V_1) + \\
&+ N^\textrm{pe}_1(V_1, V_2) \mdot \lambda_{\textrm{eCT}_{12}}(V_1, V_2))
\end{split}
\label{eqn:compound:gen}
\end{align}

Assuming a symmetric chamber with the same bias for both photo-detectors, Equation~\ref{eqn:compound:gen} simplifies as if there were only a single photo-detector with $\lambda_\textrm{oCT} = \lambda_\textrm{iCT} + \lambda_\textrm{fCT} + \lambda_\textrm{eCT}$:
\begin{align}
&N_1^\textrm{pe}(V) = \frac{\nicefrac{\bar{n}^\textrm{pe}}{2} \mdot E\mdot P_\textrm{PDE}(V)}{1 - \lambda_\textrm{oCT}(V)} = N_2^\textrm{pe}(V) \\
&\textrm{LY}_G(V)=\frac{\bar{n}^\textrm{pe} \mdot P_\textrm{PDE}(V)}{1 - (\lambda_\textrm{iCT}(V)+ \lambda_\textrm{fCT}(V) + \lambda_\textrm{eCT}(V))} 
\label{eqn:lyg}
\end{align}

\begin{figure}[b]
\centering
\includegraphics[width=\columnwidth]{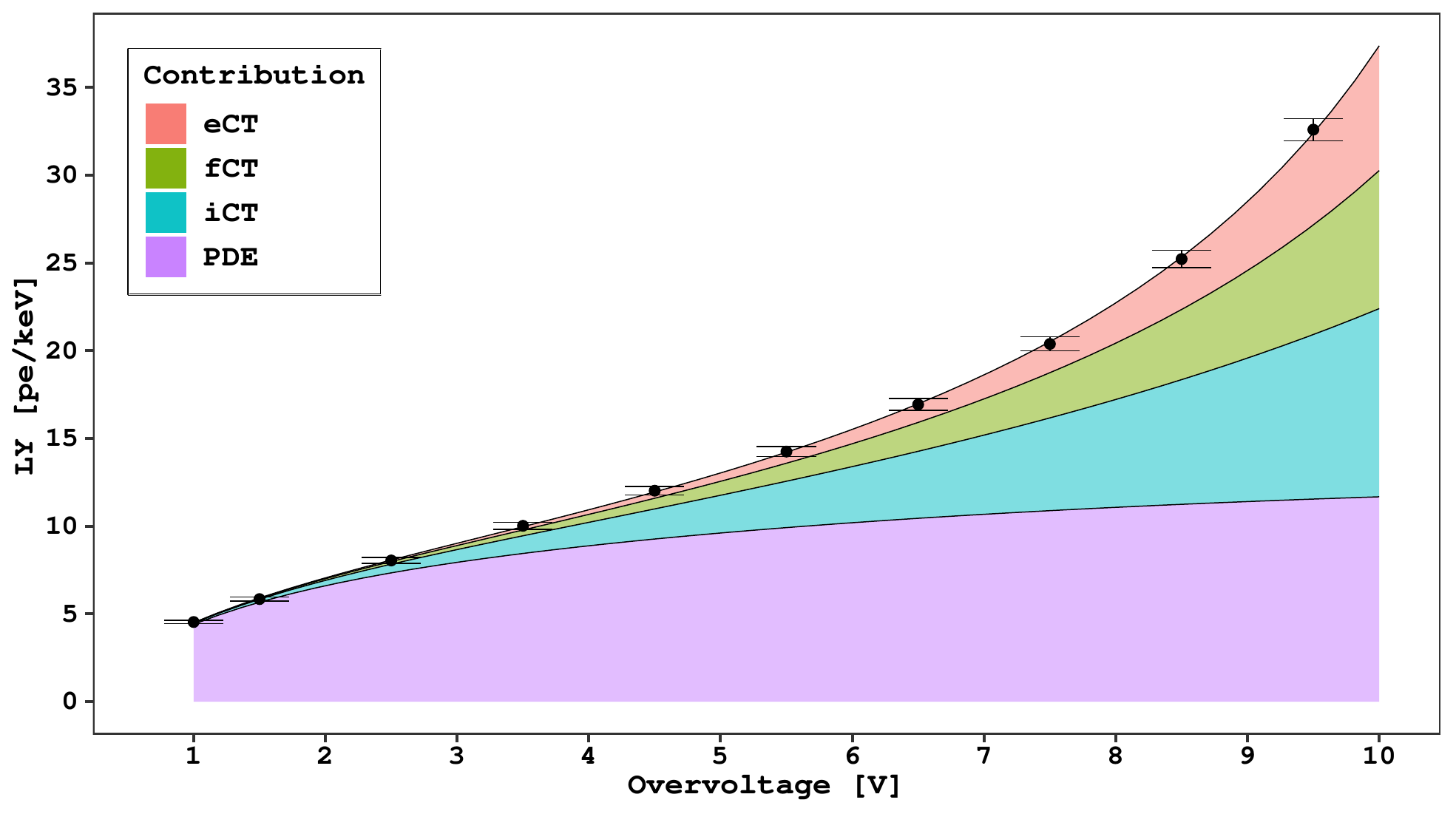}
\caption{Breakdown of the CT components from the analytical model for the measured LY$_G$ for the cubic chamber using \isotope[83\rm m]Kr, versus the over-voltage.  Each band is estimated by zeroing the acceptances in Equation~\ref{eqn:lyg} and, therefore, the eCT and fCT bands include a non-negligible fraction of iCT due to the detection of the corresponding cross-talk photons. The top black line represent the fit to the data using Equation~\ref{eqn:lyg}.}
\label{fig:breakdown}
\end{figure}

Equation~\ref{eqn:lyg} can be fit to the experimental data: it contains seven parameters $\bar{n}^\textrm{pe}$, $\zeta$, $V_h$, $V_e$, $\xi_\textrm{iCT}$, $\xi_\textrm{fCT}$, and $\xi_\textrm{eCT}$.  $V_h$ is set to the value obtained in Section~\ref{sec:ict} with a penalty in the chi-squared.  The fit to the data from the symmetric setup (equal tile biases) is only sensitive to $\lambda_\textrm{oCT}$, and not to the individual components.  We therefore set
$\xi_\textrm{iCT}$ and $\xi_\textrm{fCT}$ to the values obtained in Sections~\ref{sec:ict} and~\ref{sec:tmc}, respectively.  The results of the minimization are summarized in Table~\ref{tab:ly:fit}.  Figure~\ref{fig:breakdown} reports the fit to the gross light yield with different colors indicating the different cross-talk contributions.  Interestingly, above \SI{7.7}{OV} the contribution from the optical cross-talk exceeds the number of primary photo-electrons.

\subsection{Energy Resolution}
\label{sec:fano}

\begin{figure}[tb]
\centering
\includegraphics[width=\columnwidth]{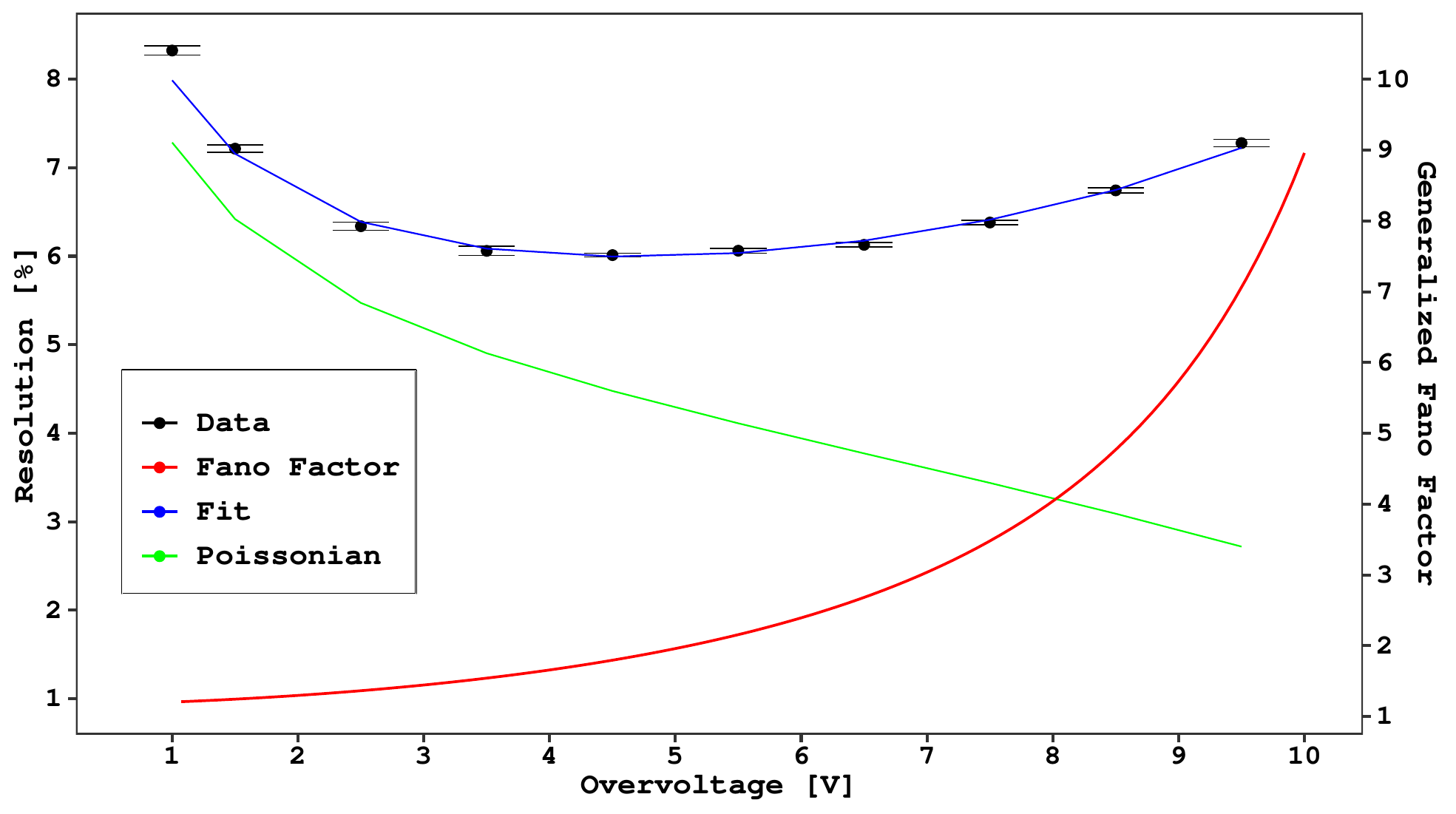}
\caption{Energy resolution of the cubic chamber using \isotope[83\rm m]Kr (data points in black).  The solid blue line represents the prediction from the resolution fit (see text).  The solid green line shows the pure Poissonian resolution and the red shows the calculated generalized Fano factor from Equation~\ref{eqn:fano2} (relative to the axis scale on the right).}
\label{fig:res}
\end{figure}

Another quantity of primary importance for particle detectors is the energy resolution.  In our case, due to iCT, the resolution diverges from the Poissonian limit as described by a larger-than-unity generalized Fano factor, introduced in Section~\ref{sec:ict}.  The presence of fCT and eCT further increase this divergence.  Analogous with Equation~\ref{eqn:ict:fano}, we define a global generalized Fano factor from cross-talk contributions as
\begin{align}
\mathcal{F}_\textrm{oCT} = \delta^*\mdot(1 - \lambda_\textrm{oCT})^\alpha
\label{eqn:fano2}
\end{align}
where $\alpha$ is the same as defined for the iCT only (Section~\ref{sec:ict}). $\delta^*$ includes the contribution from the SiPMs, plus the spread introduced by the argon scintillation~\cite{fanoAR} and TPB wavelength shifter~\cite{Francini}. 
This results in
\begin{align}
\frac{\sigma_G(E, V)}{E} = \sqrt{\frac{\mathcal{F}_\textrm{oCT}(V)}{E\mdot\textrm{LY}_G^\textrm{OV}}},
\label{eqn:sigma}
\end{align}
which can be fit to experimental data.

For the cubic chamber, with \isotope[83\rm m]Kr, we obtain a $\chi^2/n.d.f. = 12/7$ when fixing $V_h=5.4$ with a resulting $\xi_\textrm{oCT}$ (sum of the three CT acceptances) equal to
\SI{78.0\pm0.3}{\per\kilo\volt} and $\delta^*$=\num{1.157\pm0.006}.  At \SI{1}{OV} the contribution of the electronic noise is dominant.
We therefore exclude the \SI{1}{OV} data point from the fit.  Figure~\ref{fig:res} shows the experimental data and the fit prediction. 

\subsection{Global fit}
We combine the LY$_G$ and the resolution fits into a global $\chi^2$ with the goal of reducing the uncertainty on the results.  We write
\begin{align}
\frac{\sigma_G(E, V)}{E} = \sqrt{\frac{\mathcal{F}_\textrm{oCT}(V)}{E\mdot\textrm{LY}_G(V)}},
\label{eqn:sigmafit}
\end{align}
where LY$_G(V)$ is predicted by the analytical model instead of the experimental data.  The results of the $\chi^2$ minimization of Equation~\ref{eqn:sigmafit} are summarized in Table~\ref{tab:ly:fit}.

\section{Optics}
A dedicated Monte Carlo simulation software based on Geant4~\cite{ALLISON2016186} 
was developed, with the main focus on tracking the photons in the detectors under study.  The simulation includes wavelength shifting, reflection, refraction, and absorption.  Additionally, it incorporates the LAr scintillation process originating from particle interactions in LAr.  The inner surface of the active LAr volume is almost entirely covered with TPB and the wavelength-shifting efficiency of TPB is set to unity in the model. 

A full model of the system was implemented, along with the surrounding LAr buffer and the detector geometry.  Refractive indices, attenuation lengths, and surface properties of the LAr, TPB, ESR foil, fused-silica windows, and SiPMs are included with adequate approximations. Some of these parameters were taken from literature and some were measured in dedicated setups.  In particular, measurements of the wavelength-dependent reflectivity of the TPB-coated ESR foil (approx.~96\%) and of the SiPM surface (approx.~17\%) are reported in Ref.~\cite{2pac}.

To estimate the light collection efficiency of the detector under study, \SI{41.5}{\kilo\eV} electrons were generated uniformly over the active volume, simulating the energy deposition from the \isotope[83\rm m]Kr isotope.  Table~\ref{tab:mc_phkill} summarizes the fraction of visible photons absorbed by different parts of the detector: the light collection is only affected by the fraction of light absorbed by the SiPM tiles.  Figure~\ref{fig:asym} compares the simulated top-bottom asymmetry with the measured value.

\begin{table}[tb]
    \centering
    \begin{tabularx}{\columnwidth}{@{\extracolsep{\fill}}lr}
    \hline 
    {\bf Component } & {\bf Fraction (\%)}  \\
    \hline
    SiPM (collection efficiency)    &   70 \\
    Dead spaces in the SiPM tile    &   6 \\
    Reflectors \& TPB   &   11 \\
    Detector mechanics (escape)    &   13 \\
    \hline
    \end{tabularx}
    \caption{Fraction of simulated photons absorbed by the different detector components.  Absorption by SiPMs represents the light collection efficiency of the system.  Absorption by detector mechanics implies the light is escaping from the optical system of the detector.}
    \label{tab:mc_phkill}
\end{table}
\label{sec:simulation}

Several issues arise when modeling the optics, the most important of which are: (1) the nature of the optical interface between the fused-silica window and TPB not being well understood, and (2) the reflectivity of the SiPM, which has a multi-layer anti-reflective coating, is measured in air and projected to the LAr medium~\cite{2pac}.  These issues affect the aforementioned light collection efficiency and, hence, we estimate a systematic uncertainty of \SI{\pm5}{\percent} on its absolute value. 

With the collection efficiency it is possible to predict the LY of the system. Assuming that the PDE of the SiPM does not change by more than \SI{10}{\percent} with respect to the measurements at room temperature~\cite{ds20k}, and a \SI{100}{\percent} for the TPB VUV conversion efficiency~\cite{WlsReview}, the optics model predicts a net light yield of \num{11.7\pm1.3} photo-electrons/keV at \SI{6}{OV} in agreement with the value reported in Figure~\ref{fig:breakdown}.

\section{Conclusions}
In this work we parametrized the over-voltage dependence of optical cross-talk (oCT) in two small LAr chambers equipped with large SiPM arrays, using an inclusive oCT model.  
The model involving internal (iCT) and external (fCT and eCT) components well describes the observed gross light yield and energy resolution, thus providing a plausible estimate of the contributions from primary photo-electrons and different cross-talk components.  We obtained a net light yield up to \num{12\pm 1}~photo-electrons/keV, which is one of the best light yield values obtained for an experimental setup using LAr.  Despite the high net light yield, the oCT of the SiPMs becomes dominant above \SI{7.7}{OV}, significantly compromising the energy resolution of the detectors.  Since SiPMs from any vendor are affected by external cross-talk, we recommend that previous results obtained using silicon readout be re-evaluated with a similar analysis. 

Encouragingly, the over-voltage dependence of the oCT is well described, providing an effective parametrization of the response of the FBK NUV-HD-Cryo SiPMs at cryogenic temperatures.  Even if the optical model of the chamber is tuned for the argon scintillation photons (in the UV/blue) and the optical cross-talk happens at longer wavelengths, we can use the collection efficiency as a first approximation to estimate the acceptances of fCT and eCT: 
$\xi_\textrm{fCT+eCT}\simeq\num{.6}~\xi_\textrm{iCT}$.  This implies that for every 
two photons trapped in the silicon bulk (candidate for iCT), one is escaping through the front window.

\section*{Acknowledgements}
We acknowledge support from the Istituto Nazionale di Fisica Nucleare (Italy) and Laboratori Nazionali del Gran Sasso (Italy) of INFN, from NSF (US, Grant PHY-1622415 and PHY-1812540 for Princeton University), from the Royal Society UK and the Science and Technology Facilities Council (STFC), part of the United Kingdom Research and Innovation, from the European Union’s Horizon 2020 research and innovation programme under grant agreement No 952480 (DarkWave project), and from the International Research Agenda Programme AstroCeNT (MAB\allowbreak/2018\allowbreak/7) funded by the Foundation for Polish Science (FNP) from the European Regional Development Fund.
\bibliographystyle{spphys}
\bibliography{main, intro}

\end{document}